\documentclass[twocolumn]{aastex63}
\usepackage{array}
\newcolumntype{P}[1]{>{\centCas Aering\arraybackslash}p{#1}}
\usepackage{graphicx}
\usepackage{cellspace}
\usepackage{amssymb}
\usepackage{amsmath}
\usepackage{longtable}
\usepackage{tabu}
\usepackage{color}
\usepackage{etoolbox}
\usepackage{supertabular}
\usepackage{dcolumn}
\usepackage{ wasysym }

\def\Tobs{T_{\textrm{\mbox{\tiny{obs}}}}}
\def\Tcoh{T_{\textrm{\mbox{\tiny{coh}}}}}
\def\Tref{T_{\textrm{\mbox{\tiny{ref}}}}}

\def\EatH{Einstein@Home}

\def\sci#1#2{#1\times10^{#2}}

\newcommand{\avgSeg}[1]{\overline{#1}}			%% average over segments

\newcommand{\Freq}{f}
\newcommand{\fdot}{{\dot{\Freq}}}
\newcommand{\fddot}{\ddot{\Freq}}

\newcommand{\M}{\mathcal{M}}

\newcommand{\Gauss}{\mathrm{\MakeUppercase{G}}}
\newcommand{\Signal}{{\mathrm{\MakeUppercase{S}}}}
\newcommand{\Line}{{\mathrm{\MakeUppercase{L}}}}
\newcommand{\Transient}{{\mathrm{t\MakeUppercase{L}}}}

\newcommand{\NoisetL}{{\Gauss\Line\Transient}}

\providecommand{\sc}[1]{\widehat{#1}}
\renewcommand{\sc}[1]{\widehat{#1}}

\newcommand{\BSNtsc}{{\hat\beta}_{{\Signal/\NoisetL}}}	%% odds ratio O_SN
\newcommand{\BSGLtLr}{{\hat\beta}_{{\Signal/\NoisetL r}}}

\newcommand{\F}{\mathcal{F}}		%% F-statistic

\newcommand{\avF}{\avgSeg{\F}}

\newcommand{\Nseg}{{N_{\mathrm{seg}}}}

\newcommand{\paramfdotloG}{\ensuremath{-\sci{8.0}{-9}}} %{-\sci{1.981}{-9}}} 
\newcommand{\paramfddothiG}{\ensuremath{\sci{1.1}{-18}}}%{\sci{0.196}{-18}}}
\newcommand{\paramWUcputimeHours}{8} % taken from https://gitmaster..aei.uni-hannover.de/einsteinathome/project-daemons/blobs/master/CFS_S6Bucket_setup.C line 463:
\newcommand{\paramtotalWUsmillionsGlow}{2.5} %how many WU (millions)
\newcommand{\paramtotaltemplatesGlow}{\ensuremath{\sci{5.1}{16}}} %G1 ntemplates
\newcommand{\paramtotaltemplatesPPGlow}{\ensuremath{\sci{2.5}{10}}} %G1 postproc ntemplates
\newcommand\Tstrut{\rule{0pt}{2.9ex}}       % "top" strut
\newcommand\Bstrut{\rule[-1.3ex]{0pt}{0pt}} % "bottom" strut
\newcommand\TBstrut{\Tstrut\Bstrut}  

\newtoggle{fullauthorlist}
\toggletrue{fullauthorlist}
\newtoggle{endauthorlist}
\toggletrue{endauthorlist}
\shorttitle{ 
Results from an Einstein@Home search for CWs from G347.3 at low frequencies in LIGO O2 Data
}
\shortauthors{Ming et al.}
\begin{document}

\title{ 
Results from an Einstein@Home search for continuous gravitational waves from G347.3 at low frequencies in LIGO O2 data
}

\author{J. Ming}
\email{jing.ming@aei.mpg.de}
\affiliation{Max Planck Institute for Gravitational Physics (Albert Einstein Institute), Callinstrasse 38, D-30167 Hannover, Germany}
\affiliation{Leibniz Universit\"at Hannover, D-30167 Hannover, Germany}

\author{M.A. Papa}
\email{maria.alessandra.papa@aei.mpg.de}
\affiliation{Max Planck Institute for Gravitational Physics (Albert Einstein Institute), Callinstrasse 38, D-30167 Hannover, Germany}
\affiliation{University of Wisconsin Milwaukee, 3135 N Maryland Ave, Milwaukee, WI 53211, USA}
\affiliation{Leibniz Universit\"at Hannover, D-30167 Hannover, Germany}

\author{H.-B. Eggenstein}
\affiliation{Max Planck Institute for Gravitational Physics (Albert Einstein Institute), Callinstrasse 38, D-30167 Hannover, Germany}
\affiliation{Leibniz Universit\"at Hannover, D-30167 Hannover, Germany}

\author{B. Machenschalk}
\affiliation{Max Planck Institute for Gravitational Physics (Albert Einstein Institute), Callinstrasse 38, D-30167 Hannover, Germany}
\affiliation{Leibniz Universit\"at Hannover, D-30167 Hannover, Germany}

\author{B. Steltner}
\affiliation{Max Planck Institute for Gravitational Physics (Albert Einstein Institute), Callinstrasse 38, D-30167 Hannover, Germany}
\affiliation{Leibniz Universit\"at Hannover, D-30167 Hannover, Germany}
 
\author{R. Prix}
\affiliation{Max Planck Institute for Gravitational Physics (Albert Einstein Institute), Callinstrasse 38, D-30167 Hannover, Germany}
\affiliation{Leibniz Universit\"at Hannover, D-30167 Hannover, Germany}

\author{B. Allen}
\affiliation{Max Planck Institute for Gravitational Physics (Albert Einstein Institute), Callinstrasse 38, D-30167 Hannover, Germany}
\affiliation{University of Wisconsin Milwaukee, 3135 N Maryland Ave, Milwaukee, WI 53211, USA}
\affiliation{Leibniz Universit\"at Hannover, D-30167 Hannover, Germany}

\author{O. Behnke}
\affiliation{Max Planck Institute for Gravitational Physics (Albert Einstein Institute), Callinstrasse 38, D-30167 Hannover, Germany}
\affiliation{Leibniz Universit\"at Hannover, D-30167 Hannover, Germany}

\begin{abstract}
We present results of a search for periodic gravitational wave signals with frequency between 20 and 400 Hz, from the neutron star in the supernova remnant G347.3-0.5, using LIGO O2 public data. The search is deployed on the volunteer computing project Einstein@Home, with thousands of participants donating compute cycles to make this endevour possible. We find no significant signal candidate and set the most constraining upper limits to date on the amplitude of gravitational wave signals from the target, corresponding to deformations below $10^{-6}$ in a large part of the band. At the frequency of best strain sensitivity, near $166$ Hz, we set 90\%\ confidence upper limits on the gravitational wave intrinsic amplitude of $h_0^{90\%}\approx 7.0\times10^{-26}$. Over most of the frequency range our upper limits are a factor of 20 smaller than the indirect age-based upper limit.

\end{abstract}


\section{Introduction}
\label{sec:introduction}

Continuous gravitational waves are among the gravitational wave signals that have not yet been detected. 
Fast spinning neutron stars with non-axisymmetric deformations or with unstable r-modes are expected to emit continuous waves which lie in the high-sensitivity frequency range of ground-based interferometers  \citep{Owen:1998xg,Owen:2010ng,lasky_2015}. 

Although the expected waveforms are fairly simple, the search for continuous wave signals is very challenging due to their extreme weakness.  Signal-to-noise ratio (SNR) is accumulated by integrating the signal over many months, and this increases our ability to resolve different waveforms. This also means that if the signal waveform is not a priori known, many different waveforms must be searched for, and the computing cost increases very significantly. In fact, when searching a broad range of waveforms, the sensitivity of continuous wave searches is usually limited by the computing power. 

Since the Advanced LIGO \citep{2015aligo} detectors begun observations, various continuous waves searches have been carried out.   
Among them, the searches for continuous waves from known pulsars, with known spin frequency and frequency evolution, are the most sensitive and computationally inexpensive   \citep{LVC_2019_targeted,LVC_2020_targeted,Ashok:2021fnj}.
At the other extreme, there are the all-sky surveys with no prior information of frequency and sky location \citep{Dergachev:2021ujz,Dergachev:2020fli, Dergachev:2020upb,Steltner_2021, lvc_o2_as, lvc_o3_as, LIGOScientific:2021tsm, O2AS_binary}.
In-between, the directed searches target locations in the sky that are known or suspected to harbour a neutron star, albeit pulsation shave generally not been observed.  Searches of this type include the galactic centre \citep{Piccinni_o2gc,Dergachev:2019pgs}, young supernova remnants (SNRs)  \citep{Ming:2019xse, Papa_2020midth, lvc_o1_snr, Millhouse_o2_snr,Lindblom:2020rug,LIGOScientific:2021mwx}, glitching pulsars \citep{Fesik:2020tvn,LIGOScientific:2021yby} and low-mass X-ray binaries such as Scorpius X-1 \citep{Zhang_2021}.

Young neutron stars are good continuous wave candidates : an indirect upper limit can be placed on continuous gravitational wave strength that is proportional to $1/\sqrt{\tau}$, with $\tau$ being the age of the neutron star \citep{LIGOScientific:2008hqb,Zhu:2016ghk}. Fifteen young supernova remnants have been identified in our Galaxy that could host a young neutron star and potentially be promising targets. Recent searches probe emission from {\it{all}} of these over a broad range of waveforms  \citep{Lindblom:2020rug,LIGOScientific:2021mwx}. 

An alternative approach is to identify the most promising targets and concentrate the search efforts on these. 
In \cite{Ming:2015jla} we propose an optimisation scheme to decide how to spend the computing budget in such a way to maximise the detection probability. With a computing budget of a few months on the Einstein@Home volunteer computing project, the indication is to search for emission from the neutron star in the SNRs Vela Jr. (G266.2-1.2), Cassiopeia A (G111.7-2.1) and G347.3 (G347.3-0.5). We carried out searches using O1 data, and O2 data for follow-ups, and set the most constraining  upper limits on gravitational wave emission from these sources with those data  \citep{Ming:2019xse,Papa_2020midth}.

In  \cite{Papa_2020midth}, we also found a sub-threshold candidate at around  369 Hz. Gravitational wave follow-ups were not completely conclusive and we found no evidence of pulsations from searches of archival X-ray data to validate this candidate, but the X-ray searches had limited sensitivity. \cite{LIGOScientific:2021mwx} did not find this candidate in the first half of O3 data, however the sensitivity of \cite{LIGOScientific:2021mwx} is lower than that of our original search. We thus prioritize a deep search for G347.3 in the O2 data below 400 Hz. This paper presents results from such a search.

The paper is organised as follows: in Section \ref{sec:target} we review the astrophysical target and the model gravitational waveform. After a brief description of the data in Section Section \ref{sec:Data},  in Section \ref{sec:search} we describe the search. The results follow in Section \ref{sec:results}, where we explain how the $h_0^{90\%}$ intrinsic continuous gravitational wave amplitude upper limits are derived. These are also recast as upper limits on the star's ellipticity and r-mode saturation. We conclude with a discussion of the results, comparing and contrasting with existing literature in Section \ref{sec:conclusions}.

\section{The Target}
\label{sec:target}

\subsection{G347.3-0.5}
\label{sec:G347.3}
The supernova remnant G347.3 is suggested to be the remnant of the AD393 ``guest star" \citep{1997A&A...318L..59W}. We therefore assume an age of 1600 years, albeit this estimate is not completely uncontroversial \citep{2012AJ....143...27F}. Using XMM data, \cite{2004A&A...427..199C} estimate its distance to be around 1.3 kpc. The position of the central compact object in the G347.3 SNR is given with sub-arcsecond accuracy in \cite{2008A&A...484..457M}, based on Chandra data.
Among the SNRs in our galaxy, G347.3 is one of the most interesting directed search targets because of its relatively young age and close distance \citep{Ming:2015jla}. %As we did in \cite{Ming:2017anf}, the optimisation scheme is applied for the directed searches for CW signals from galactic SNRs using O2 data. The optimisation scheme results a specific search setup for the source G347.3 at the frequency band from 20 to 400 Hz. The expected $h_0$ upper limit of this search would be the best up to date. 

In the deep CW search for G347.3 in O1 data \citep{Papa_2020midth}, we find an interesting candidate at around 369 Hz. The spin-down energy loss from the candidate parameters yields an unusually high value,  $1.6\times10^{40}$ erg/s, which exceeds the most energetic Crab pulsar's $\dot{E} = 4.6\times10^{38}$ erg/s and J0537-6910's $\dot{E} = 4.9\times10^{38}$ erg/s. %Although this candidate is not confirmed by X-ray searches in \cite{Papa_2020midth}, it is very interesting to perform a full run of CW search for G347.3 in O2 data particularly at low frequency range 20-400 Hz.

\subsection{The Signal}
\label{sec:signal}

We assume a standard IT2 continuous gravitational wave signal  \citep{Dergachev:2020upb} produced by asymmetric rotating neutron stars which, in the detector data, has a form \citep{Jaranowski:1998qm}:
\begin{equation}
h(t)=F_+(t)h_+(t)+F_{\times}(t)h_{\times}(t),
\label{eq:signal}
\end{equation}
where $F_+(t)$ and $F_\times(t)$ are the antenna pattern functions of the detector for the two gravitational wave polarizations ``+" and ``$\times$".
They depend on the sky position of the source (defined by the right ascension $\alpha$ and
declination $\delta$), and the orientation $\psi$ of the wave-frame with respect to the detector frame.
 $F_+(t)$ and $F_\times(t)$ are periodic time functions with a period of one sidereal day, because the detector rotates with the Earth.

The phase $\Phi(t)$ of the signal at solar system barycenter (SSB) frame has the form:
\begin{multline}
\label{eq:phiSSB}
\Phi(\tau_{\mathrm{SSB}}) = \Phi_0 + 2\pi [ f (\tau_{\mathrm{SSB}}-{\tau_0}_{\mathrm{SSB}})  +
\\ {1\over 2} \dot{f} (\tau_{\mathrm{SSB}}-{\tau_0}_{\mathrm{SSB}})^2 + {1\over 6} \ddot{f} (\tau_{\mathrm{SSB}}-{\tau_0}_{\mathrm{SSB}})^3 ], 
\end{multline}
where $f$ is the signal frequency and $\tau_{\mathrm{SSB}}$ is the arrival time of the GW front at the SSB frame.

\section{The data}
\label{sec:Data} 

The LIGO O2 public data  \citep{o2_data,Abbott:2021boh} is used in this search.
 The data is from the two observatories in the USA, one in Hanford (Washington State) and the other in Livingston (Louisiana). The data used in this search is between GPS time
1167983370 (Jan 09 2017) and 1187731774 (Aug 25 2017). Short Fourier transforms (SFTs) of data segments 1800 seconds long  \citep{SFTs} are created as customary for \EatH \, searches.

Calibration lines, the mains power lines and some other spurious noise due to the LIGO laser beam jitter are removed in the publicly released O2 data  \citep{Davis_2019}. Additionally we remove loud short-duration glitches with the gating procedure described in  \cite{Steltner:2021qjy} and substitute Gaussian noise in the frequency domain in disturbed bins. This is a standard procedure in Einstein@Home searches.

\section{The Search}
\label{sec:search}

We use a  ``stack-slide"  type of search based on the GCT (Global correlation transform) method  \citep{PletschAllen,Pletsch:2008,Pletsch:2010}. 
The data is partitioned in $\Nseg$ segments and each segment spans a duration $\Tcoh$. The data of both detectors from each segment $i$ is searched with a maximum likelihood coherent method to construct the detection statistic, $\F$-statistic  \citep{Cutler:2005hc}. 
The statistics $\F_i$ from the coherent searches of the different segments are summed, and the value of the core detection statistic $\avF$ is obtained: 
\begin{equation}
\label{eq:avF}
\avF:={1\over\Nseg} \sum_{i=1}^{\Nseg} \F_i.
\end{equation}

In Gaussian noise $\Nseg\times 2\avF$ follows a chi-squared distribution with $4\Nseg$ degrees of freedom, and a non-centrality parameter $\rho^2$. If a  signal is present, $\rho^2$ is proportional to $h_0^2\Tobs \over S_h$, where $S_h$ is the strain power spectral density of the noise at the frequency of the signal, and $h_0$ is the signal intrinsic amplitude at Earth  \citep{Jaranowski:1998qm}.

The data in reality is not Gaussian and despite the removal of loud glitches and lines, some coherent disturbances persist. The $\avF$ can be effected by these coherent disturbances and present increased values. In order to identify occurrences of this, a line robust detection statistic $\BSNtsc$  \citep{Keitel:2013wga,Keitel:2015ova} is computed. 
This statistic is the log of a Bayesian odds ratio that tests the signal hypothesis versus an extended noise hypothesis. 
The noise model of this statistic not only includes Gaussian noise, but also coherent single-detector signals. 
%In the appendix \ref{sec:parameters} we provide the values of the parameters used for this detection statistic.
The \EatH \, results from this search are ranked according to $\BSNtsc$, such that the top-list contains fewer candidates which are affected by the coherent disturbances.

The search set-up, i.e. the coherent baseline $\Tcoh$, the template grid spacings and the search ranges are all derived from the optimisation procedure.

We search for signal-waveforms with frequency and frequency-derivatives as follows:
\begin{equation}
\label{eq:Priors}
	\begin{cases}
	20 ~\mathrm{Hz} \le f  \le 400 ~\mathrm{Hz}\\
	-f/ \tau\,  \le   \dot{f}  \le 0\,~\mathrm{Hz/s}\\
	0\,\mathrm{Hz/s}^2 \leq  \ddot{f} \leq ~7\dot{|f|}_{\textrm{max}}^2/f = 7 {f/\tau^2}, 
	\end{cases}
\end{equation}
where $\tau=1600 \,\textrm{years}$. The ranges for $\dot{f}$ and $\ddot{f}$ correspond to different breaking index $n$ values, namely 2 and 7.
In the $\dot{f}$ equation the $n=2$ is used to encompass the broadest range of $\dot{f}$ values. In the $\ddot{f}$  equation  $n=7$ is used to encompass all  astrophysical scenarios including the phase evolution purely due to GW emission ($n=5$) and r-mode oscillations ($n=7$). 
At 400 Hz, the  $\dot{f}$ extends down to \paramfdotloG  ~ Hz/s and the $\ddot{f}$ range up to \paramfddothiG~ Hz/s$^2$.

\begin{figure}[h!tbp]
  \includegraphics[width=\columnwidth]{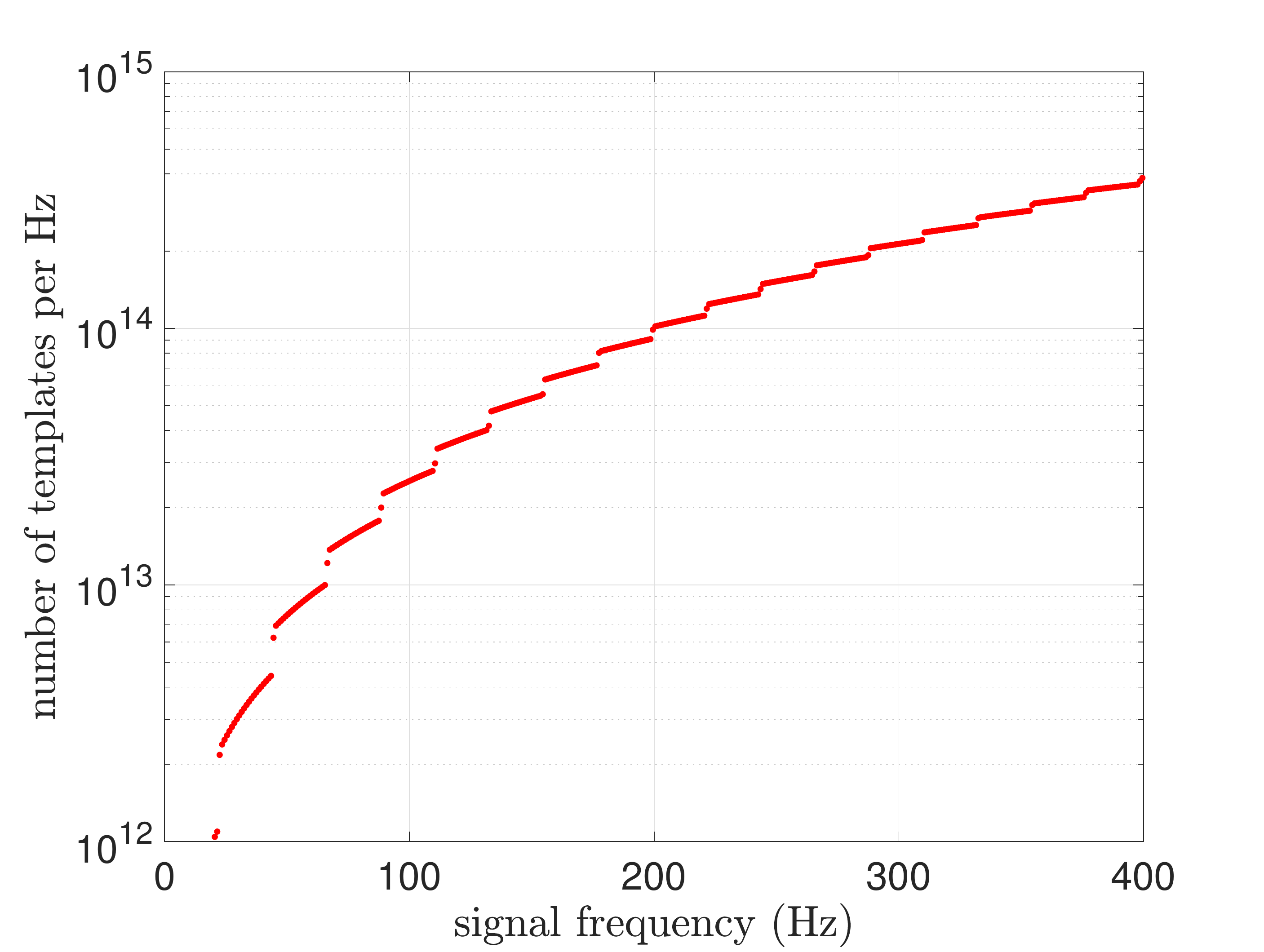}
\caption{Number of templates searched in 1-Hz bands as a function of signal frequency. }
%$N_f \times N_{\dot{f}} \sim 1.3\times 10^{9}$, where $N_f$ and $N_{\dot{f}}$ are the number of $f$ and $\dot{f}$ templates searched in 50\,mHz bands. The total number of templates searched between 20 and 100 Hz is {\paramtotaltemplates}. % #N f*fdot templates includest gamma refine
%}  
\label{fig:HowManyTemplates}
\end{figure}

The grid spacings in frequency and spin-downs are constant over these search ranges and are given in Table \ref{tab:GridSpacings}. The number of searched templates per 1 Hz band increases as the frequency increases, as Eq.~\ref{eq:Priors} shows.
Fig. \ref{fig:HowManyTemplates} shows the number of templates searched in 1-Hz bands as a function of frequency. 

%In frequency the same grid spacing is used in the coherent searches over the segments and in the incoherent summing step. In $\dot{f}$ and $\ddot{f}$, the coherent step and the incoherent step require different grid spacings. The final template grid, on which the $\avF$ is evaluated, on a finer template grid than the one used for the single segment $\F_i$. 

%The grid spacings for the coherent step in  $\dot{f}$ and $\ddot{f}$ are  $\delta{\dot{f_c}}, \delta{\ddot{f_c}}$ respectively. For the incoherent step the grid spacings are finer by factors $\gamma_1$ and $\gamma_2$.
%Followed by the optimisation scheme, these spacing values are obtained. 
%These parameters are given in Table \ref{tab:GridSpacings}.

%%%%
%%%%%
\begin{table*}[ht]
%\begin{table}[t]
\begin{centering}
\begin{tabular}{|c|c|c|c|c|c|}
\hline
\hline
  \multicolumn{6}{|c|}{ SEARCH SET-UP} \\
\hline
\hline
 \TBstrut$\Tcoh=1080$ hr &  $\Nseg=5$ & $\delta f=1.3 \times 10^{-7}$ Hz & $\delta {\dot{f}}=1.5\times 10^{-14}$ Hz/s & $\delta {\ddot{f}}=1.2\times 10^{-20}$ Hz/s$^2$ &  $\Tref= 1177858472.0\footnote{Barycentric Dynamical Time in GPS seconds}$  \\
 \hline
\hline  
%$(\alpha,\delta) ~{\text{sky position}}$ & $ $ \\
% \hline
\end{tabular}
\caption{Spacings on the signal parameters used for the templates in the search. }
\label{tab:GridSpacings}
%\end{minipage}
\end{centering}
\end{table*}

%%%%

The search is performed on the \EatH~volunteer computing project. \EatH~is built on the BOINC (Berkeley Open Infrastructure for Network Computing) architecture~ \citep{Boinc2,Boinc3} which uses the idle time on volunteer computers to tackle scientific problems such as this, that require large amounts of computing power.

Overall we search $\approx$ {\paramtotaltemplatesGlow} templates, utilizing \EatH \, for several weeks. The work-load is split in work-units, sized to keep the average volunteer host busy for {\paramWUcputimeHours} hours. The whole search task is split into about {\paramtotalWUsmillionsGlow} million work-units. 
%This requires 10,000  (A)-type CPUs\footnote{(A)-type CPUs are CPUs of Einstein@Home hosts in the fast performance catalog. For a same computing task,  a (A)-type CPU is averagely 3 times faster than  a (B)-type CPU. More details can be found at Section 3 of  \citep{Ming:2017anf}.  } of \EatHs running for 1 month,  not including redundancy for cross-validation.  
Only information from the most promising 10000 results from each work-unit is communicated back to the central \EatH~server.

%due to the different coherent time-baselines used for the different targets, due to the different mismatch of the grids and because of the different search ranges due to the different ages of the targets. Fig. \ref{fig:HowManyTemplates} shows the number of templates searched as a function of frequency for the three targets. 

%\mapcomment{MAP edited up to here}

\section{Results}
\label{sec:results}

%\section{Post-processing of the search result}

After the \EatH~server has received all search results, the post-processing begins.
In total we have ${\paramtotalWUsmillionsGlow} ~{\textrm{million work-units}}\times 10\,000 ~{\textrm{results returned per work-unit}}\approx \paramtotaltemplatesPPGlow$ search results. Each result is identified by the template-waveform parameters ($f,\fdot,\fddot$) and by the detection statistics values.

%The first post-processing step is called {\bf{``banding"}}:   
%Each work-unit searches 50 mHz in frequency and the entire $\ddot{f}$ range at every frequency. The $\dot{f}$ range is tuned to keep the computing-time constant. The banding procedure is a sort of inverse procedure of dividing up the computing load among work-units, by splitting up the searched $\dot{f}$ range among different work-units. In the banding step, all results the same 50 mHz band are assembled together. \mapcomment{if we did not remove any disturbed band, we do not need to talk about the banding.}

%Using the plots with statistics, the procedure of identification of disturbed bands can be started.
%\jingcomment{I don't remember we did this. For now I leave this part blank}
%\mapcomment{Please check if you have any "holes" in the bands that you took the candidates from.}
%\jingcomment{I have checked. There are five candidates(max of 0.5 Hz)  from holes. However, I think this does not mean these candidates can not be used for the upper limit simulation. For example, one candidate has frequency 350.250 hz at the ref time. In the h data, let's say there is  very narrow gaussian noise  band from  350.250+-1e-4 Hz. In the L data there is no gaussian noise  band. To compute the 2Fr of this candidate, the data used spans half Hz. So 2$\times$ half Hz vs 2e-4 Hz, 99.98\% of the data used to compute the 2F of this candidate is the real data. The candidate should still be considered as a valid candidate which is used to set the Th in the simulation of upper limit. }
With a parameter-space clustering procedure we identify the most interesting results  \citep{Steltner_2021,Singh:2017kss,Beheshtipour:2020zhb,Beheshtipour:2020nko}.  We refer to these as ``candidates". 
%that are close to each other Due to the same root-cause, some part of the candidates are clustered and close by each other in parameter space.
%The procedure "Clustering" is applied to pick out the candidate which has the highest $\BSGLtLr$. 
%The Clustering procedure let us to keep the candidates which are most representative and excludes the ones less interesting. 
%As it is used in  \citep{Steltner_2021}, a new clustering method is applied in this paper. 
%Compared with the adaptive clustering procedure  \citep{Singh:2017kss}, this new clustering method results a lower false dismissal of signals at fixed false alarm rate.
%\jingcomment{More details for the clustering needs to be filled by HB. However, I am not sure if we should give more details of clustering, because even without clustering, we have the same upper limit. }
We consider the top 1 million candidates, corresponding to a detection statistic threshold $\BSGLtLr=1.948$. 
%Note that, in this paper, the further investigation results of these top 1 million candidates is not presented. It will be released in the next paper. This paper presents the results of $h_0^{90\%}$ with the assumption that non of the 1 million candidates can be vetoed or proved to be a signal in the follow-up searches. 

\begin{figure}[h!tbp]
  \includegraphics[width=\columnwidth]{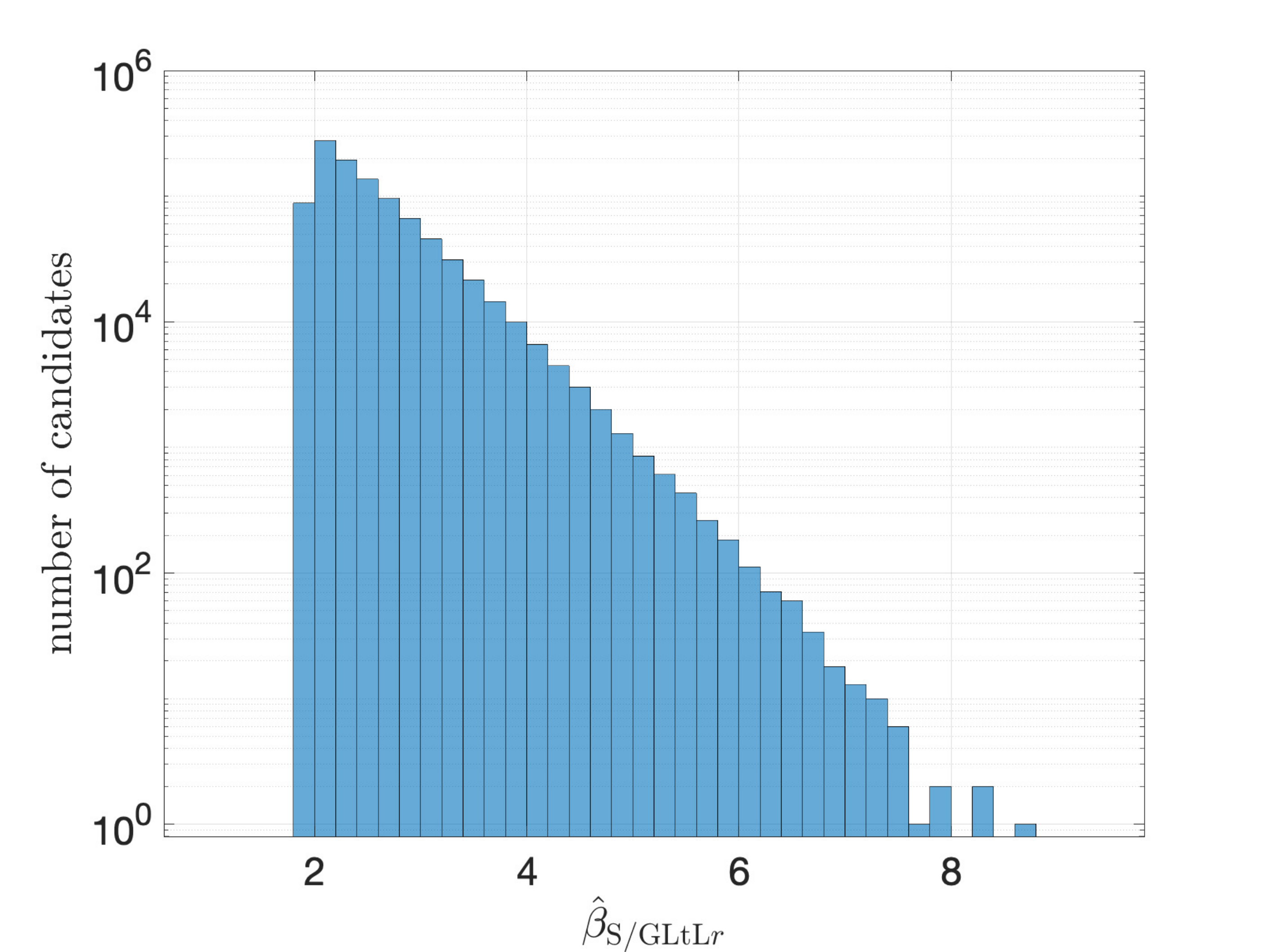}
   \includegraphics[width=\columnwidth]{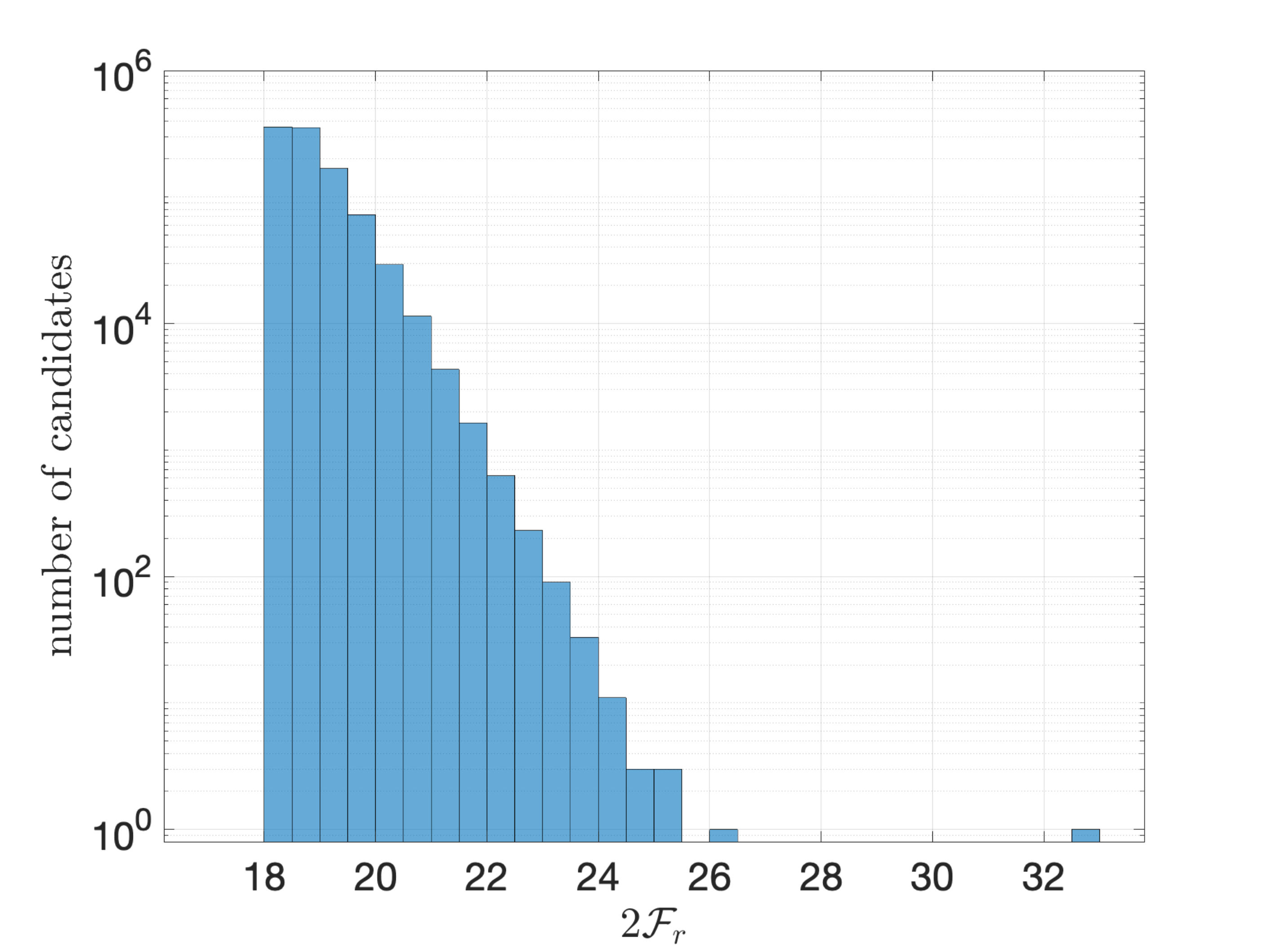}
\caption{Distribution of the detection statistics  $\BSGLtLr$ (top) and $2\F_r$ (bottom) of the top 1 million candidates ranked according to $\BSGLtLr$, which is the line- and transient-line -robust statistic.}
%$N_f \times N_{\dot{f}} \sim 1.3\times 10^{9}$, where $N_f$ and $N_{\dot{f}}$ are the number of $f$ and $\dot{f}$ templates searched in 50\,mHz bands. The total number of templates searched between 20 and 100 Hz is {\paramtotaltemplates}. % #N f*fdot templates includest gamma refine
%}  
\label{fig:hist_BS}
\end{figure}

%\section{Results}
%\label{sec:results}
%32.9 at stage 0
%%54.6 at stage 1, with twice Tcoh
%%
The distribution of the detection statistic  $\BSGLtLr$ and $2\F_r$ for these candidates is shown in Figure \ref{fig:hist_BS}. We use $\BSGLtLr$ to rank our candidates but also show $2\F_r$ because its distribution in Gaussian noise is known. A detectable signal would look like an obvious outlier in both distributions. In Figures \ref{fig:hist_BS} we instead see an outlier in the $2\F_r$ distribution but not in the $\BSGLtLr$ distribution. This is an indication that a coherence in one of the two detectors is causing the high value of $2\F_r$. In particular the $2\F_r$ outlier has a value of 32.9, whereas its $\BSGLtLr=2.0$ which is in 5th percentile of  lowest values. We follow-up this candidate with a semi-coherent search with $\Tcoh = 2760$ hr. The most significant result  $2\F_r=54.6$ falls short of what one would expect from a signal: none of the over thousand signals tested showed this small increase in signal-to-noise ratio. After excluding this candidate,  Figures \ref{fig:hist_BS} shows no significant signal candidate in either $2\F_r$ or $\BSGLtLr$ .

%\mapcomment{MAP looked at paper until here}

\subsection{Upper limits}
We determine the smallest $h_0$ that would have produced a detection statistic as high as the most significant measured in every half Hz band. We assume the source to be at the position of our target, the spindown to be in the target range and the frequency varying in each half Hz. 
%Since we do not know the inclination angle $\iota$, we assume a uniform $\cos\iota$ distribution and marginalise over this. 
We set the confidence level at 90\%, meaning that 90\% of the signals in the considered range with an amplitude at the upper limit value $h_0^{90\%}$, would yield a value of the detection statistic larger than the loudest search result from that parameter range. We use the $\BSGLtLr$ as our reference statistic, since it is our ranking statistic. 

In each half Hz band, 200 simulated signals with a fixed value of the intrinsic amplitude $h_0$  are added to the real detector data. The data is then processed as the data that was searched, i.e. it is gated and line-cleaned. 

The parameters of simulated signals, the  frequency, inclination angle $\cos\iota$, polarization $\psi$ and initial phase values, are uniformly randomly distributed in their respective ranges. The spin-down values, $\dot{f}$ and $\ddot{f}$, are log-uniformly randomly distributed in their respective ranges. 

A search is performed to recover each injection with the same grid and set-up as the original Einstein@Home search. The search is more limited than the original search to save computations, and covers the parameter space neighbouring the fake signal. The fake signal is counted as recovered if the $\BSGLtLr$ from the search is higher than the maximum $\BSGLtLr$ from the Einstein@Home results in the same half Hz band. 

This whole procedure is repeated for various values of $h_0$. For each value of $h_0$,  the fraction of detected injections is determined in this way and varying $h_0$ the confidence $C(h_0)$ curve is constructed. We use a fit with a sigmoid of the form: 
\begin{equation}
C(h_0)={1\over{1+\exp({{{\textrm{a}}-h_0}\over{\textrm{b}}})}},
\label{eq:sigmoidFit}
\end{equation}
and from it we read-off the $h_0$ amplitude that corresponds to 90\% confidence, our upper limit value.

\begin{figure}[h!tbp]
   \includegraphics[width=\columnwidth]{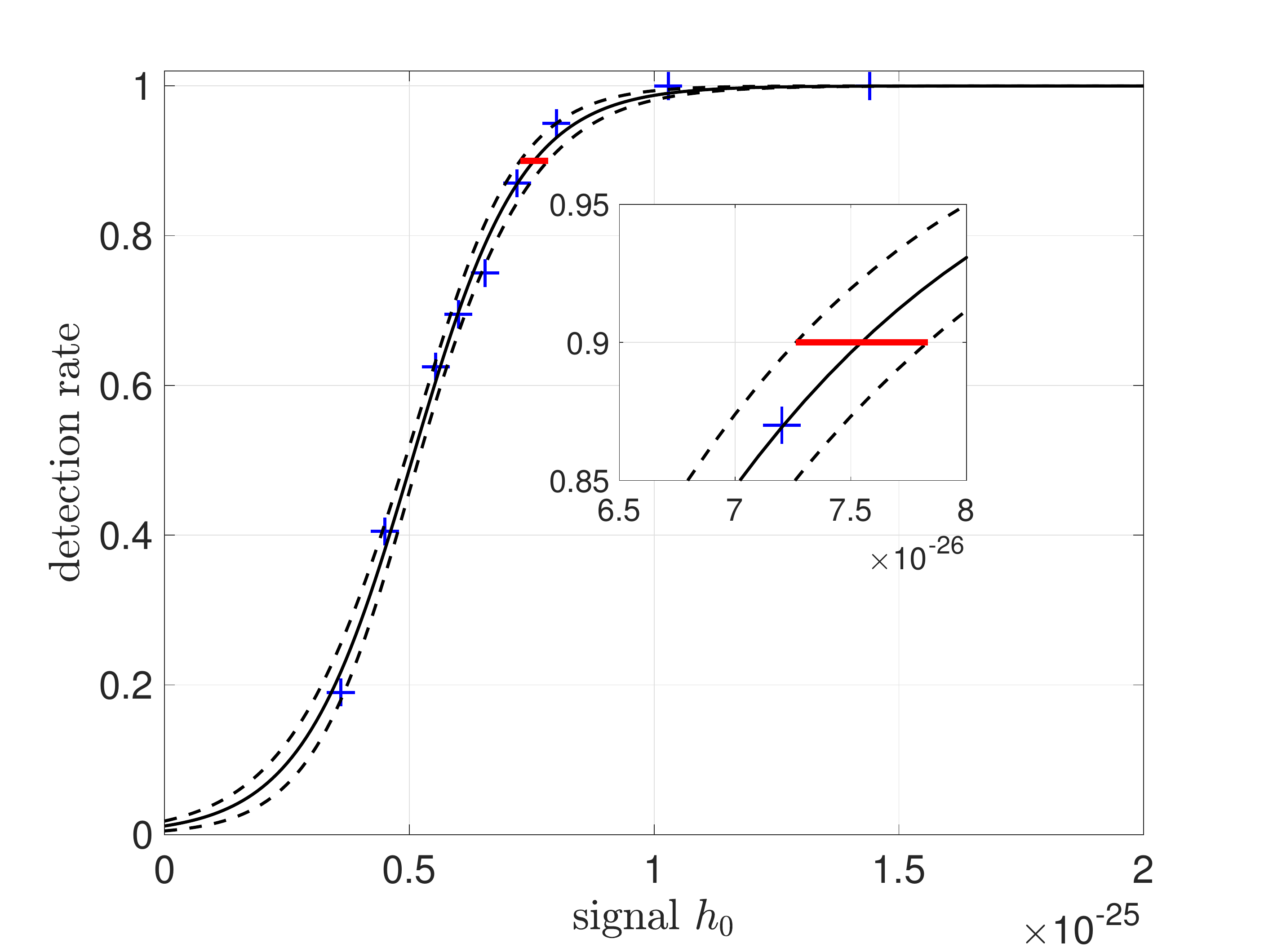}
\caption{Blue crosses: measured detection efficiency $C(h_0)$ from search-and-recovery Monte Carlos in the frequency band 149 to 149.5 Hz. The solid line is the best fit and the dashed lines represent 95\% confidence intervals on the fit. The red line marks the 90\% detection rate, with the uncertainties introduced by this fitting procedure 4\%. The inset shows a zoom around the 90\% confidence level.}
\label{fig:exampleSigmoid}
\end{figure}

The Matlab nonlinear regression prediction confidence intervals routine {\tt{nlpredci}} is used to yield the best-fit for ${\textrm{a}}$ and ${\textrm{b}}$ values and the covariance matrix. This  covariance matrix can be used to compute the 95\% credible interval on the fit of $h_0^{90\%}$.
Figure \ref{fig:exampleSigmoid} shows the sigmoid curve fitting for the 149-149.5 Hz band, as  a representative example of the results obtained with this procedure. 
The best fit for  $h_0^{90\%}$ in this band is $7.5\times10^{-26}$. The uncertainties introduced by this procedure are less than 4\%. The total uncertainty in the upper limit is the sum of the fitting procedure uncertainty and the calibration uncertainties. We conservatively use 5\% as the calibration uncertainty \citep{PhysRevD.96.102001}. 

\begin{figure*}%[h!tbp]
   \includegraphics[width=0.85\textwidth]{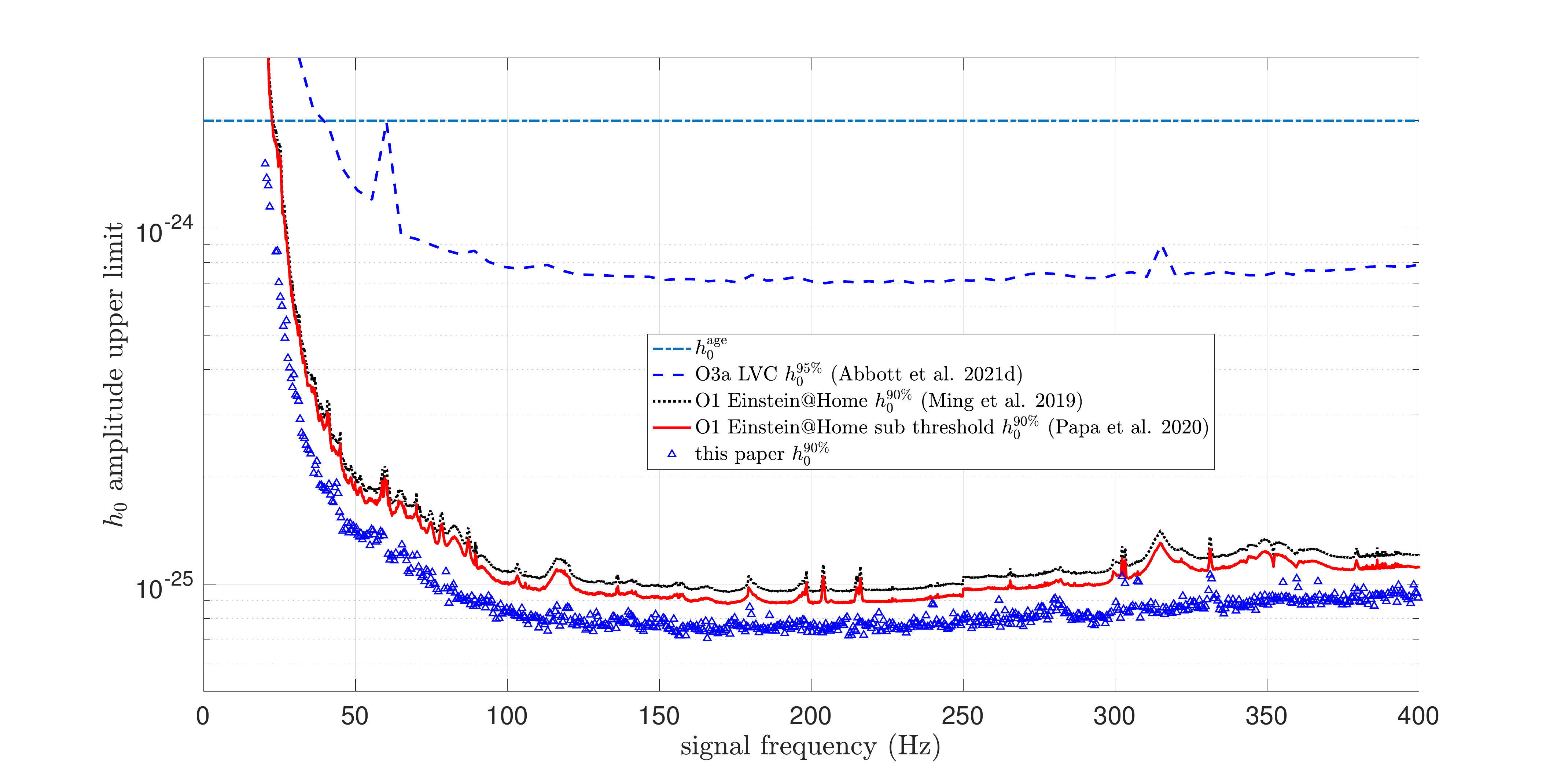}
\caption{90\% confidence upper limits on the gravitational wave amplitude of continuous gravitational wave signals from G347.3 for signals with frequency between  20 to 400 Hz. 
The lower blue triangles are the  results of this search and we compare them with results from previous searches: The blue dots are the upper limits from the LVC search of the O3a \citep{LIGOScientific:2021mwx};  The black dots are Einstein@Home results from O1 data \citep{Ming:2019xse} and red solid line the sub-threshold search \citep{Papa_2020midth}. 
}  
\label{fig:ULs}
\end{figure*}

The $h_0^{90\%}$ upper limits are shown in Figure \ref{fig:ULs} and provided in machine readable format at \cite{AEIULurl}. 

In nineteen half-Hz bands we do not set an upper limit; correspondingly in the upper limit files we have 741 entries rather than 760. The cleaning procedure substitutes disturbed frequency-domain data with Gaussian noise in order to avoid further spectral contamination from ``leakage" in the search results. Those bands are consistently cleaned in the upper-limit Monte Carlos after a signal is injected, so it may happen that most of the injected signal is  removed. When that happens, no matter how loud the signal is, the detection efficiency does not increase. In these bands the 90\% detection rate level cannot be reached and we do not set any upper limit. This reflects the fact that, even if we had a signal there, because of the cleaning procedure, we could not detect it. 

In other bands the cleaning procedure partly or completely removes some of the signals, depending on their frequency.  So, in order to produce a detection statistic value above a given threshold, statistically, a louder signal is required than in nearby bands that are not cleaned. In those bands the upper limit is higher than what it would be if the data had not been cleaned. 
For example, $h_0^{90\%}$ of the band 331 - 331.5 Hz is is about 15\% larger than the $h_0^{90\%}$ of the neighbor half Hz bands. In this bands 8\% of the data is Gaussian noise data. 

\subsection{Upper limits on the astrophysical parameters}

\begin{figure}[h!tbp]
   \includegraphics[width=\columnwidth]{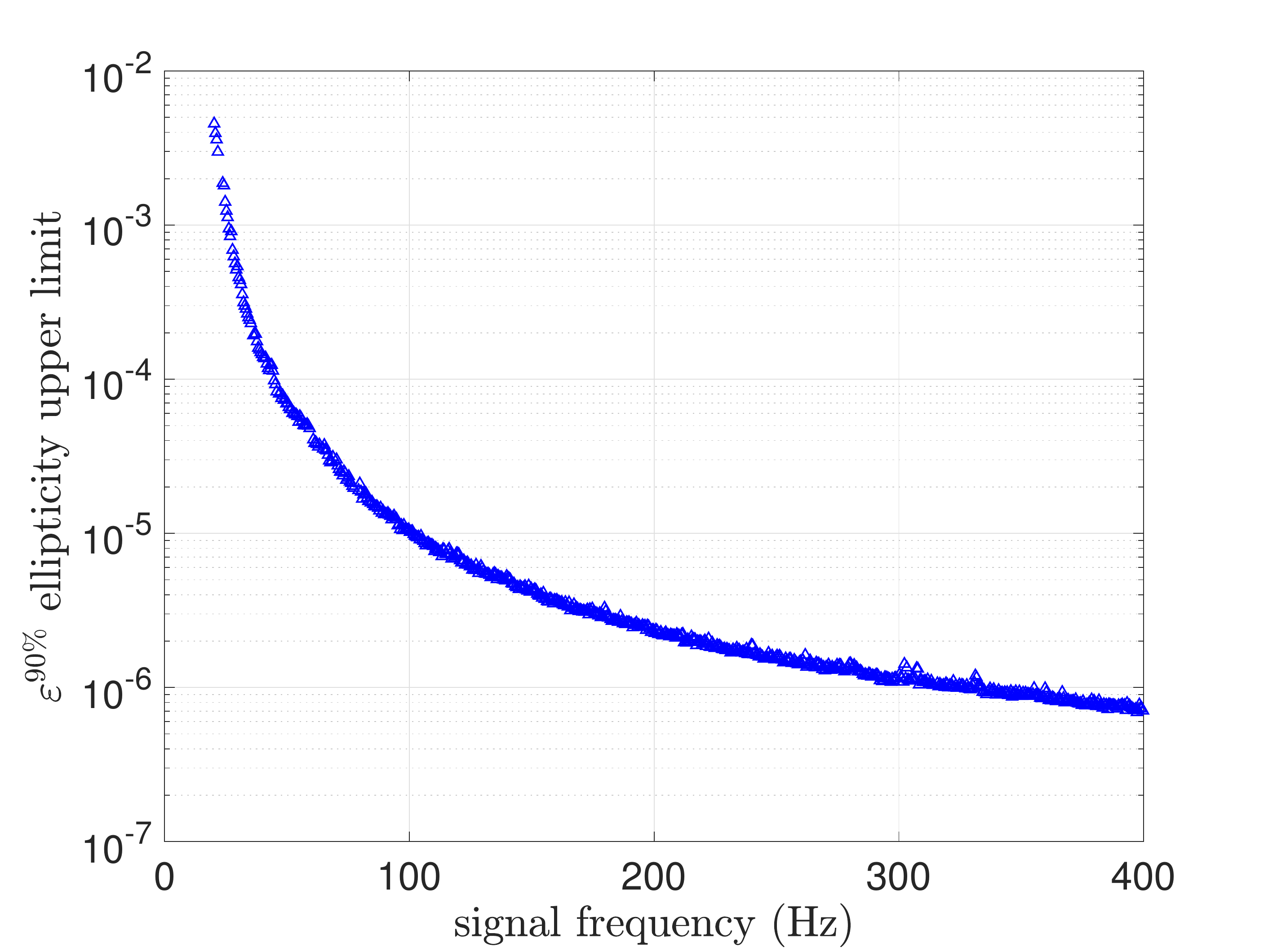}
\caption{ Upper limits on the equatorial ellipticity of G347.3. We assume a distance of 1.3 kpc.}
\label{fig:epsilonULs}
\end{figure}

The $h_0$ upper limits can be converted in constraints on the equatorial ellipticity $\varepsilon$ of the neutron star at a distance $D$ and at frequency $f$ \citep{zimmermann:1979}: 
\begin{equation}
\varepsilon = {{c^4}\over {4\pi^2 G}}{{h_0 D}\over {I f^2}},
\label{eq:epsilon}
\end{equation}
where $c$ is the speed of light, $G$ is the gravitational constant and $I$ the principal moment of inertia of the star. 
Assuming a fiducial value of the principal moment of inertia of $10^{38} \textrm{kg m}^2$ and $D=1.3\,\mathrm{kpc}$, we convert $h_0^{90\%}(f)$  into  upper limits on the ellipticity of the source G347.3. These are shown in Figure~\ref{fig:epsilonULs}.

%Above  $\approx320$ Hz, the $\varepsilon^{90\%}$ is lower than $10^{-6}$. At 400 Hz , the highest frequency of this search, the $\varepsilon^{90\%}$ reaches its lowest upper limit of  $6.9\times10^{-7}$. 

\begin{figure}[h!tbp]
   \includegraphics[width=\columnwidth]{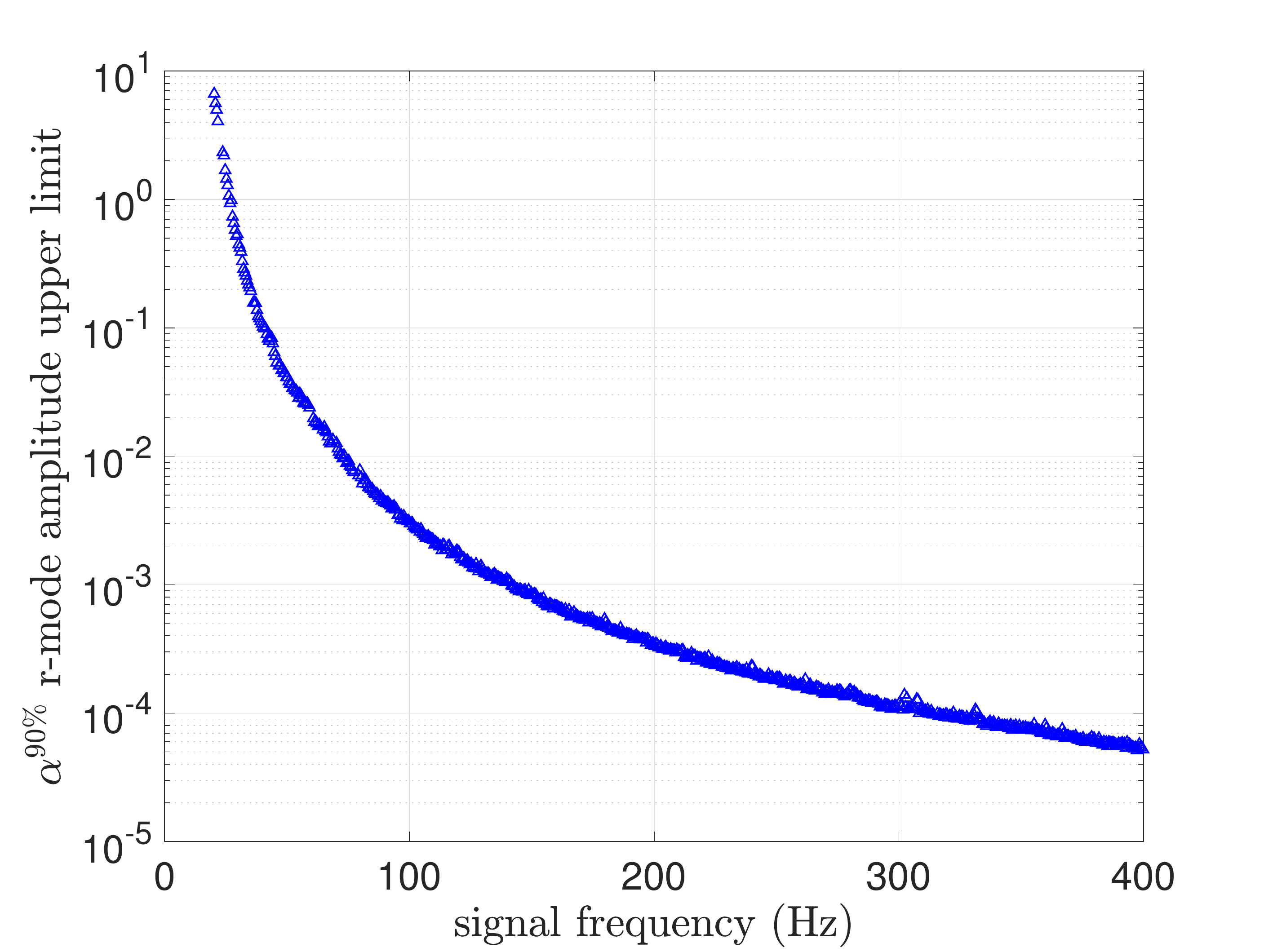}
\caption{Upper limits on the r-mode amplitude.}  
\label{fig:alphaULs}
\end{figure}

R-mode oscillations of a spinning neutron stars also produce continuous gravitational waves. The  amplitude $h_0$ for a signal with frequency $f$, from a source at a distance $D$, depend on the r-mode amplitude $\alpha$ as follows \citep{Owen:2010ng}:
\begin{equation}
\alpha = 0.028 \left( {h_0\over{10^{-24}}}\right )\left ( {D\over{1~\textrm{kpc}}}\right ) \left ({{\textrm{100~Hz}}\over f} \right )^3.
\label{eq:rmodes}
\end{equation}
Our $h_0^{90\%}$ upper limits can then be recast as upper limits on the r-mode amplitude. The result is shown in Figure \ref{fig:alphaULs}. 
%Starting from  $\approx310$ Hz, the $\alpha^{90\%}$ is blow $1\times10^{-4}$.  At the highest frequency 400 Hz of this search, the $\alpha^{90\%}$ reaches its lowest upper limit $5.1\times10^{-5}$.% \citep{Owen:2010ng}

\section{Conclusions}
\label{sec:conclusions}

In this paper we present results from the most sensitive search to date for continuous gravitational wave emission from the supernova remnant G347.3-0.5 in the frequency range 20-400 Hz and the broadest first and second frequency-derivative range. Electro-magnetic pulsations have not been detected from this object, and the direct observation of continuous gravitational emission would provide the first gravitational wave pulsar timing solution.

We prioritize this target with respect to other SNRs because of a sub-threshold candidate from a previous search. We do not find a signal.  

We constrain the amplitude of continuous gravitational wave emission at a level which is more than a factor of 20 smaller than the indirect age-based limit over most of the frequency range. The  most constraining intrinsic gravitational wave amplitude upper limit is $7.0\times 10^{-26}$ near 166 Hz. This result improves over our O1 result \citep{Ming:2019xse}  and over the extensive sub-threshold O1 search \citep{Papa_2020midth}.  It is also more constraining than the recent search result of  \cite{LIGOScientific:2021mwx} that uses the significantly more sensitive O3a data. In fairness we note however  \cite{LIGOScientific:2021mwx} search a broader frequency range and their search uses a technique that is more robust to possible deviations of the signal from the IT-{\it{n}} model.

Recast in terms of equatorial ellipticity of the neutron star, our results constrain it below $10^{-6}$ at frequencies higher than $\approx$ 320 Hz reaching bounds of $6.9\times 10^{-7}$ at 400 Hz. This is a physically plausible value of neutron star deformation \citep{McDanielJohnsonOwen,Gittins:2020cvx,Gittins:2021zpv}. Such limit is not matched in \cite{LIGOScientific:2021mwx} even at 2000 Hz. 

Our spindown range is high enough to allow for braking indexes as high as 7, encompassing r-mode emission. Our null result can then constrain the r-mode amplitude and does so at a level below $10^{-4}$ at frequencies higher than $\approx$ 310 Hz. This is also a physically possible value \citep{Haskell:2015iia}.

This is the first O2 public-data Einstein@Home search for continuous gravitational waves from SNRs and probes a physically interesting range of source parameters. Building on this, future searches will extend the parameters space and/or include more targets and/or more data, pushing further in interesting territory.

\section{Acknowledgments}

We gratefully acknowledge the support of the many thousands of Einstein@Home volunteers who made this search possible. \\
We acknowledge support from the Max Planck Society for Projects QPQ10003 and QPQ10004, and the NSF grant 1816904.\\ 
A lot of post-processing is run on the ATLAS cluster at AEI Hannover. We thank Carsten Aulbert and Henning Fehrmann for their support. \\
We would like to thank the instrument-scientist and engineers of LIGO whose amazing work has produced detectors capable of probing gravitational waves so incredibly small.\\
This research has made use of data, software and/or web tools obtained from the Gravitational Wave Open Science Center (https://www.gw-openscience.org/ ), a service of LIGO Laboratory, the LIGO Scientific Collaboration and the Virgo Collaboration. LIGO Laboratory and Advanced LIGO are funded by the United States National Science Foundation (NSF) as well as the Science and Technology Facilities Council (STFC) of the United Kingdom, the Max-Planck-Society (MPS), and the State of Niedersachsen/Germany for support of the construction of Advanced LIGO and construction and operation of the GEO600 detector. Additional support for Advanced LIGO was provided by the Australian Research Council. Virgo is funded, through the European Gravitational Observatory (EGO), by the French Centre National de Recherche Scientifique (CNRS), the Italian Istituto Nazionale di Fisica Nucleare (INFN) and the Dutch Nikhef, with contributions by institutions from Belgium, Germany, Greece, Hungary, Ireland, Japan, Monaco, Poland, Portugal, Spain.

% references
\newpage

%%%% Useful bibliography (not Phys Rev D format), with titles %%%%%
%\if 0 1
%\begin{thebibliography}{99}

%\input{paperBib.tex}
%\end{thebibliography}
\bibliography{paperBibApJ}
\bibliographystyle{aasjournal}

%\appendix
\newpage

\end{document}